\documentclass{optica-article}

\journal{opticajournal} 

\articletype{Research Article}

\usepackage{lineno}
\usepackage{subcaption}
\usepackage{physics}
\usepackage{multirow}
\begin{document}

\title{Low-Loss Polarization-Maintaining Router for Single and Entangled Photons at a Telecom Wavelength}

\author{Pengfei Wang,\authormark{1} Soyoung Baek,\authormark{1, 2} Masahiro Yabuno,\authormark{1,3} Shigehito Miki,\authormark{3} Hirotaka Terai,\authormark{3} and Fumihiro Kaneda\authormark{1,4,*}}

\address{
\authormark{1}Graduate School of Science, Department of Physics, Tohoku University, 6-3 Aramaki aza Aoba, Aoba-ku, Sendai 980-8578, Japan\\
\authormark{2}Research Institute of Electrical Communication, Tohoku University, 2-1-1 Katahira, Aoba-ku, Sendai 980-8577, Japan\\
\authormark{3}Advanced ICT Research Institute, National Institute of Information and Communications Technology, 588-2 Iwaoka, Nishi-ku, Kobe 652-2492, Japan\\
\authormark{4}Precursory Research for Embryonic Science and Technology (PRESTO), Japan Science and Technology Agency (JST), Kawaguchi 332-0012, Japan
}

\email{\authormark{*}fumihiro.kaneda.b6@tohoku.ac.jp} 


\begin{abstract*}
Photon polarization serves as an essential quantum information carrier in quantum information and measurement applications. 
Routing of arbitrarily polarized single photons and polarization-entangled photons is a crucial technology for scaling up quantum information applications. 
Here, we demonstrate a low-loss, noiseless, polarization-maintaining routing of arbitrarily polarized single photons and, crucially, multi-photon entangled states where the entanglement is encoded in orthogonal polarization bases, at the telecom L-band.
Our interferometer-based router is constructed by optics with a low angle of incidence and cross-aligned electro-optic crystals, achieving the polarization-maintaining operation with a minimal number of optical components.
We demonstrate the routing of arbitrarily-polarized heralded single photons with a 0.057 dB (1.3\%) loss, a $>$ 22 dB switching extinction ratio, and $>$ 99\% polarization process fidelity to ideal identity operation. 
Moreover, the high-quality router achieves the routing of two-photon N00N-type entangled states with a highly maintained interference visibility of $\approx$ 97\%. 
The demonstrated router scheme preserving multi-photon polarization state paves the way toward polarization-encoded photonic quantum network as well as
multi-photon entanglement synthesis via spatial- and time-multiplexing techniques.
\end{abstract*}

\section{Introduction}
Optical networking is an essential technology in various types of telecommunications. Routing of optical signals is a fundamental operation in both classical and quantum information and communication technologies. While classical optical communication systems handle intense optical pulses, quantum information applications need to transmit and route a single photon, which is fragile against loss and recoverable with only probabilistic schemes \cite{oqc, errorcorrection,clone}. Thus, a photonic router for single photons requires low-loss, low-noise, and high-speed operation without disturbing their quantum states except for spatial modes. 

Quantum information can be encoded into a variety of optical degrees of freedom \cite{Polarization1, Polarization2, Time, Frequency, Space}. In particular, polarization is widely used as a qubit, and thus, routing photons with preserved polarization states is essential to realize a polarization-encoded quantum network. 
Such a polarization-maintaining router can also be a key component of an all-optical quantum memory \cite{kanedaoptica, kanedaSA}, which enhances the generation rate of heralded single photons via temporal multiplexing. 
Moreover, a polarization-maintaining router needs to maintain \textit{quantum entanglement} of input photons, enabling entangled-state time multiplexing \cite{entanglemulti}, active feedforward control \cite{LArozema}, quantum channel multiplexers \cite{LeeNPJQ75_2022}, and quantum repeaters \cite{repeater1, repeater2} to realize large-scale quantum computing, long-distance quantum communication, and photonic quantum-state synthesis \cite{MccuskerPRL2009}.
The router should also be operated at a telecom band, where low-loss fiber networks, sources of high-indistinguishability single photons \cite{kanedaSA} and high-quality entangled photons \cite{BSM1,BSM2}, and high-efficiency single-photon detectors \cite{SNSPD} have already been available. 
Previous works have demonstrated the routing of single photons, although the simultaneous achievement of low-loss and polarization-maintaining operation remains a challenge: A bulk electro-optic modulator (EOM) achieved low-loss operation ($\approx$ 0.04 dB)\cite{kanedaSA}, but the birefringence of EO crystals makes it unavailable for arbitrarily polarized single photons and polarization-entangled photons. 
Polarization-maintaining bulk EO routers have been demonstrated by introducing additional birefringence compensator crystals \cite{antonZ} and polarization-path mode conversion optics \cite{entanglemulti,interface}, which accompany extra loss and beam wavefront distortion degrading switching extinction ratio (SER). 
All-optical approaches utilizing the nonlinear Kerr effect in optical fiber have achieved the ultrafast routing of both single photons \cite{THz} and photonic entanglement \cite{ultrafast} at a $>$ GHz switching bandwidth. 
However, these schemes suffer from insertion loss in fiber-based optical components and require careful separation of signal photons from noise photons. 
Other methods involving waveguided EO switches \cite{waveguideEOM,sagnac} and intra-cavity difference-frequency generation \cite{cavity} have significant losses induced by waveguide coupling and intra-cavity propagation. 
Previously, we proposed and experimentally demonstrated a low-loss and polarization-maintaining router at a wavelength of 780 nm using a custom EOM with cross-aligned EO crystals\cite{JJAP}. 
Despite the potential for routing arbitrarily polarized single photons, the demonstrated router is not available in a telecom band, and its operational stability was restricted to only a few minutes. 
This short operational duration is primarily attributed to the router's fragile configuration, which is highly susceptible to environmental disturbances. 
Specifically, the large transverse separation between the two interferometer arms and the asymmetric placement of an EOM on only one arm contribute to this instability.

Here, we demonstrate a low-loss, polarization-maintaining router that switches an optical path of arbitrarily polarized heralded single photons and polarization-photon-number-entangled photons at the telecom L-band. 
Our scheme utilizes a semi-common-path Mach-Zehnder interferometer (MZI) with a low angle of incidence and the push-pull operation of birefringence-compensated EOMs, enabling the $2 \times 2$ polarization-maintaining switching with a minimal number of optical components for low insertion loss. 
With optimized optical components at the L-band, our router achieves an insertion loss of 0.057 dB (1.3\%), a $\approx$ 3 ns rise/fall time, and a $>$ 22 dB SER. 
We show that arbitrarily polarized heralded single photons are routed with $>$ 99\% quantum process fidelity to an ideal identity operation for all input and output optical paths. 
Moreover, we demonstrate the routing of polarization-mode N00N-type entangled states with a highly maintained two-photon interference visibility ($\approx$ 97\%). 
To the best of our knowledge,
this is the first demonstration of actively switching optical paths of multi-photon entanglement, which is encoded into orthogonally polarized photon-number states. 
The result shows the strong potential of our router scheme for realizing polarization-maintaining all-optical quantum memories and multi-channel routers, which are crucial technologies in polarization-encoded quantum networks. 
The operation wavelength at the telecom band ensures the router's compatibility with the most developed photonic facilities, such as low-loss fiber networks, high-quality photon sources, and high-efficiency single-photon detectors.

\section{Methods}
\begin{figure}[ht!]
  \centering
  \includegraphics[width=\columnwidth]{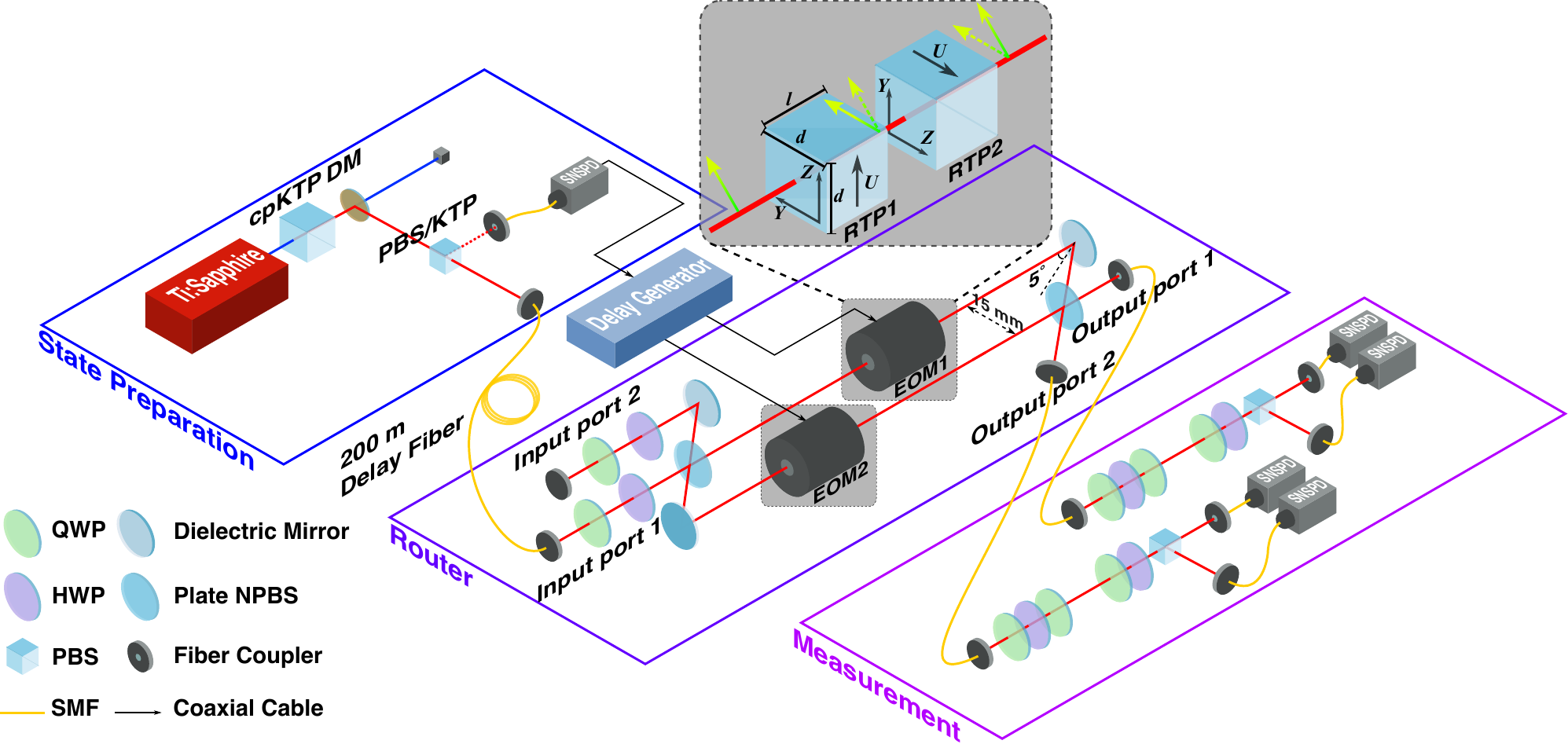}
  \caption{
  Schematic diagram of our polarization-maintaining photonic router. For stable routing operation, we employed the semi-common-path configuration of an MZI with the transverse distance between the two arms of 15 mm. The push-pull operation enabled by an EOM in each arm reduces the applied voltage by half and also improves the thermal stability of the MZI. An angle of incidence to optics component is nearly normal ($5^{\circ}$) so that they are operated independently of polarization. Note that the figure is not to scale. Inset: Polarization-maintaining EOM.  Two rubidium titanyl phosphate (RTP) crystals with orthogonally oriented crystallographic axes are placed in series. An electric field is applied to each RTP crystal along the crystallographic $Z$-axis direction. Thus, the static and EO birefringence is compensated for each other, enabling an identical phase shift to arbitrary input polarization states. QWP, quarter-wave plate; HWP, half-wave plate; PBS, polarizing beam splitter; SMF, single-mode optical fiber; NPBS, non-polarizing beam splitter; DM, dichroic mirror; cpKTP, custom-poled potassium titanyl phosphate crystal; SNSPD, superconducting nanowire single-photon detector.}
  \label{setup}
\end{figure}

Figure 1 illustrates the optical setup. The MZI-based router contained two EOMs to control the phase difference between two interferometer arms. By adjusting the phase difference, photons from an arbitrary input port can be routed to an arbitrary output port. To achieve polarization-maintaining routing, each of the EOM consisted of two cross-axis aligned rubidium titanyl phosphate (RTP) crystals (RTP1 and RTP2), as shown in the inset of Fig. 1. The RTP crystals were placed with their crystallographic $Y$ and $Z$ axes parallel or perpendicular to horizontal ($H$) and vertical ($V$) polarizations of input photons. The phase shift $\phi_{kj}$ of the $k$ ($= H, V$) polarized light given by RTP$j$ ($= 1, 2$) with applied voltage along the $Z$-axis direction can be calculated as \cite{yariv}
\begin{equation}
\phi_{V1}=\frac{2\pi}{\lambda}l_1(n_z-\frac{1}{2}r_{33}n_z^3\frac{U_1}{d_1}),~\phi_{V2}=\frac{2\pi}{\lambda}l_2(n_y-\frac{1}{2}r_{23}n_y^3\frac{U_2}{d_2}),
\end{equation}
\begin{equation}
\phi_{H1}=\frac{2\pi}{\lambda}l_1(n_y-\frac{1}{2}r_{23}n_y^3\frac{U_1}{d_1}),~\phi_{H2}=\frac{2\pi}{\lambda}l_2(n_z-\frac{1}{2}r_{33}n_z^3\frac{U_2}{d_2}),
\end{equation}
where \(\lambda\) denotes the wavelength of the incident light; \(l_j\) and \(d_j\) are the length and width of the RTP crystals ($l_1=l_2$ = 10 mm and $d_1=d_2$ = 3 mm in our experiment); \(n_y\) (\(n_z\)) and \(r_{23}\) (\(r_{33}\)) are the (static) refractive index and EO coefficient for crystallographic \(Y\) (\(Z\)) direction, respectively; \(U_j\) is the voltage applied to the crystal. For identical applied fields to the two crystals, i.e., \(U_1/d_1 = U_2/d_2\), one can achieve the same phase shift \(\phi_{V1}+\phi_{V2}=\phi_{H1}+\phi_{H2}\) for orthogonal polarization states. Thus, the static and EO birefringence of the two crystals was compensated for each other, enabling a polarization-maintaining phase shift according to the applied electric field. Note that by changing the polarity of the voltage applied to one of the crystals, the EOM can serve as a thermally compensated Pockels cell to switch polarization states. In the router, one EOM induced a positive phase shift, whereas the other a negative phase shift. This push-pull operation reduced the applied voltage to each crystal by half and improved the thermal stability of the MZI compared to the case of using a single EOM in one arm. To mitigate the polarization-dependent phase shift and transmission/reflection (on S- and P-polarizations), the other optical components were used with a nearly normal angle of incidence ($5^{\circ}$). 
As a result, \textit{all} constituent components in the router intrinsically maintained the polarization of single- and multi-photon states and their entanglement. This intrinsic polarization preservation enabled a novel compensator-free router configuration with a minimal number of optical components, each of which can cause absorption and reflection loss and wavefront distortion. Furthermore, the interferometer was constructed as a semi-common-path configuration with a 15-mm transverse displacement between the two optical paths. This achieved the stability of interference with a contrast of $> 20$ dB in a laboratory environment for 4 h (see Appendix I). In contrast, the previous prototype router \cite{JJAP}, with its 60-mm transverse distance and asymmetric placement of an EOM, could maintain 20 dB contrast for only a few minutes.

To characterize the routing of single- and entangled-photon states, a spontaneous parametric downconversion (SPDC) photon source was used,\cite{cpKTP} as depicted in Fig. 1. A Ti: Sapphire laser (with a repetition rate of 76 MHz and a central wavelength of 785 nm) was used as a pump source. A custom-poled potassium titanyl phosphate (cpKTP) crystal satisfied a Type-II quasi-phase matching condition, generating colinear single-photon pairs with a central wavelength of 1570 nm, a bandwidth of 3 nm, and a pulse width of 1.6 ps. The state of produced photon pairs was in the state $|1\rangle_H |1\rangle_V$, where $|n\rangle_k$ ($k = H, V$) is the $k$-polarized $n$-photon Fock state. Thanks to the engineered group-velocity-matching condition and the tailored spatial nonlinearity in the cpKTP crystal, the produced photons have high spectral indistinguishability ($>$ 98\%). For the generation of heralded single photons, a polarizing beam splitter (PBS) was used so that an $H$-polarized photon was heralded by the detection of a twin $V$-polarized photon by a superconducting nanowire single-photon detector (SNSPD). The SNSPD detection signal was used to trigger the EOMs to synchronize their operation with the arrival time of a heralded single photon. At the input port of the router, a half-wave plate (HWP) and a quarter-wave plate (QWP) were used to prepare an arbitrary input polarization state for heralded single photons. To generate a polarization-mode N00N-type entanglement state, the PBS was replaced by a KTP crystal (half the length of the cpKTP crystal) to compensate for the group delay of the $H$- and $V$-polarized photons. The elimination of the temporal distinguishability transformed the photon-pair state into $\frac{1}{\sqrt2}(|0\rangle_D |2\rangle_A-|2\rangle_D |0\rangle_A)$, in which a two-photon Fock state had either diagonal ($D$) or anti-diagonal ($A$) polarization states. 
Output photons from the router were analyzed by arbitrary polarization projection (using QWP, HWP, and PBS) and coincidence measurement (performed by SNSPDs and a time tagger). 

\section{Results}
\begin{figure}[t!]
  \centering
      \includegraphics[trim={310pt 205pt 315pt 210pt},clip,width=\linewidth]{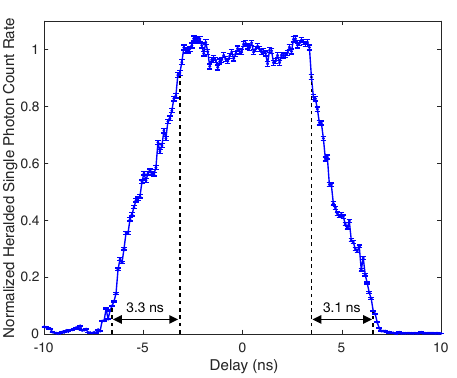}
  \caption{Normalized heralded single photon count rate versus delay time of trigger signal to the EOMs. Rise and fall times (10\% to 90\% of signal intensity transition time) of 3.3 and 3.1 ns are observed. The uncertainty is estimated from Poissonian photon counting statistics.}
  \label{speed}
  \end{figure}

Our constructed router was first characterized by classical measurements using a laser source at the telecom C-band and L-band. The measured free-space insertion loss of our router (from input beam splitter to output beam splitter) is 0.057 dB in the range of 1550-1570 nm. The degradation of the transmission is mainly due to the insertion loss by EOM1 (0.017 dB) and EOM2 (0.035 dB), resulting in an average loss of 0.026 dB, while other optics (two mirrors and two beam splitters) share an average loss of 0.031 dB. 
Our plate beam splitters are custom products (from OptoSigma) achieving a 50:50 splitting ratio at an angle of incidence of $5^{\circ}$ with beam splitter coating on one side and anti-reflection coating on the other side. The mirrors have dielectric high-reflectance coatings (ThorLabs E04) at the near-normal angle of incidence. We note that such a low insertion loss is achieved through our novel design of router configuration that has a minimal number of optical components. A beam from each output port of the router is coupled to a single-mode optical fiber with a coupling efficiency of 91\% (i.e., 0.41 dB loss). The semi-common-path configuration achieved a stable interference output for more than 4 h, as discussed in Appendix I. The router is operated at a 1 MHz repetition rate, which is determined by the high-voltage drivers. The half-wave voltage to switch the output path is $U_{\pi} = 960$ V. 

We then performed the comprehensive characterization of our router using the heralded single photons. We measured routing rise and fall times, which determine the minimum time delay of two successive pulses to be routed independently. Figure 2 shows the normalized heralded single-photon count rate on the output port 1 versus the switching time delay. We apply $U_{\pi}$ to the EOMs for 10 ns and scan the delay of trigger signals. The observed rise and fall times (in 10\% to 90\% signal transition) are 3.3 and 3.1 ns, respectively. Those are as fast as a standard Pockels cell and sufficient to resolve the arrival time of our heralded single photons pumped at a period of 13.2 ns. Small plateaus observed in the middle of the rise and fall edges may be due to the slight mismatch of the delays and temporal responses of the two EOMs.

\begin{figure}[t!]
  \centering
      \includegraphics[trim={310pt 205pt 315pt 210pt},clip,width=\linewidth]{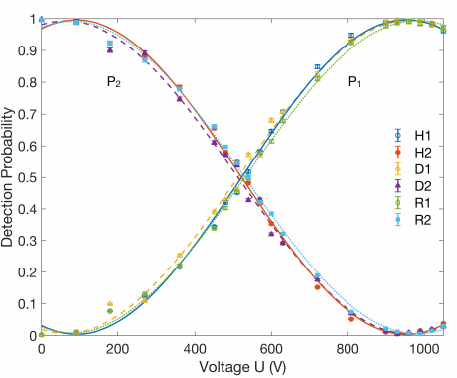}
  \caption{Relative photon detection probability at the two output ports ($P_1$ and $P_2$) versus applied voltage to the EOMs $U$ for horizontal ($H$, circles), diagonal ($D$, triangles), and right-circular ($R$, squares) input polarization states. The blank and solid symbols show the results at the output ports 1 and 2, respectively; for example, the result with horizontal input and the output port 1 ($H1$) is shown as blank circles. The solid, dashed, and dotted curves are sinusoidal data fittings for $H$, $D$, and $R$ input polarization states, respectively. The error bars are estimated by Poissonian photon counting statistics. }
  \label{switchingcurve}
\end{figure}

\begin{table}[t!]
  \centering
  \caption{SERs ($E$) and interference visibilities ($V_1$) for both output ports with $H$, $D$, and $R$ input polarization states. The uncertainty is estimated by Poissonian photon counting statistics.}
    \begin{tabular}{cccccc}
    \hline
    Input polarization & \multicolumn{2}{c}{Output port 1 ($U = U_{\pi}$)} & \multicolumn{2}{c}{Output port 2 ($U = 0$)} \\
     & $E$ (dB) & $V_1$ & $E$  (dB) & $V_1$ \\
     \hline
    $H$ & 22.2(8) & 98.80(22)\% & 25.5(10) & 99.44(14)\% \\
    $D$ & 23.3(10) & 99.07(20)\% & 28.3(16) & 99.70(12)\% \\
    $R$ & 22.6(8) & 98.90(22)\% & 25.4(12) & 99.43(16)\% \\
    \hline
    \end{tabular}
\end{table}
A polarization-maintaining router should direct single photons to an arbitrary output port, irrespective of their polarization states. 
Hence, we characterized the router’s switching extinction ratio (SER) and interference visibility of the heralded single photons with $H$, $D$, and right-circular ($R$) input polarization states, which are eigenstates of mutually unbiased Pauli bases.
Figure 3 shows the relative detection probability $P_i$ of heralded single photons as a function of applied voltage $U$ to the EOMs. Here, $P_i = S_i/( S_1 +S_2)$ is obtained using the heralded single-photon count rate $S_i$ at the output port $i$. We observe that $P_1 (P_2)$ is modulated from $0(1)$ to $1(0)$ as applied voltage, demonstrating the full tunability of the splitting ratio between two output ports. Table 1 shows the calculated SER $E$ and interference visibility $V_1$, which are calculated as:
\begin{equation}
  E = \frac{\max(S_1,S_2)}{\min(S_1,S_2)}; V_1 = \frac{\max(S_1,S_2)-\min(S_1,S_2)}{\max(S_1,S_2)+\min(S_1,S_2)},
\end{equation}
for the output port 1 at $U = U_{\pi}$ and the output port 2 at $U = 0$, respectively. The observed $E >$ 20 dB and $V_1 \approx$ 99\% for all input polarization states demonstrate our router’s capability of routing an arbitrary input polarization state with consistent performance. The slight ($<$ 1\%) degradation in SERs and interference visibilities for the output port 1 is caused by the spatial mode distortion and the polarization rotation by the EOMs operated at the high repetition rate (1 MHz) with high applied voltage ($U_{\pi} = 960$ V): we observed $V_1 >$ 99.8\% and $E >$ 30 dB when the EOMs are not operated.
Note that the high SER is achieved by the router's noiseless operation: 
the noise photon probability during the 10-ns activation of the EOMs (without single-photon input) is observed to be negligibly small compared to the dark count probability of SNSPDs ($\approx 10^{-6}$). This noiseless operation is also crucial in the observation of the polarization-maintaining routing of single and entangled photons, as will be demonstrated.

\begin{figure}[t!]
  \centering
      \includegraphics[trim={195pt 195pt 205pt 185pt},clip,width=\linewidth]{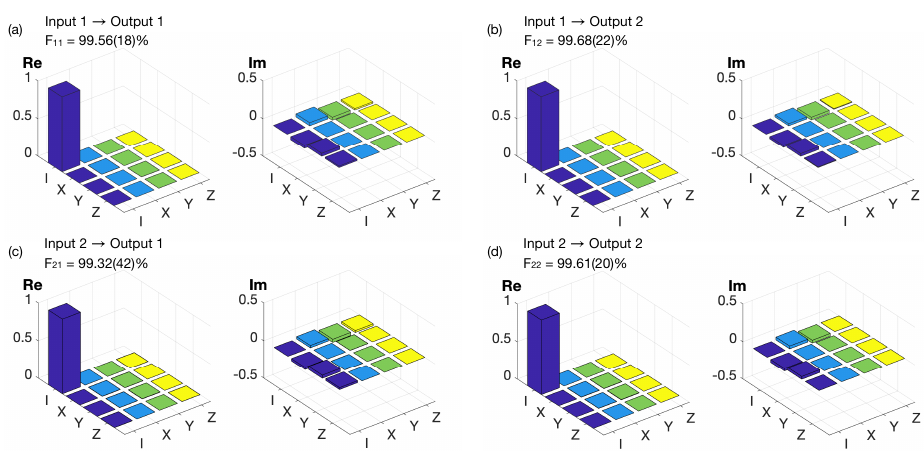}
  \caption{Reconstructed process matrix $\chi_\mathrm{R}$ in Pauli basis for the input port $i$ and the output port $j$ ($i,j = 1,2$). $F_{ij}$ denotes the process fidelity to ideal identity operation. The uncertainties of $F_{ij}$ are obtained by the standard deviations of ten independent tomographic measurement datasets. 
  }
  \label{process}
\end{figure}

To characterize the router’s polarization-maintaining operation, we performed quantum process tomography \cite{process} of the routing process, which determines the (unwanted) operation to an input single-photon polarization state. The polarization process of the router is decomposed into the Pauli basis \(\sigma_i, (i= I, X, Y, Z)\) and characterized as a process matrix $\chi_{\textrm{R}}$ with its matrix elements $\chi_{\textrm{R}ij}$:
\begin{equation}
  \rho_{\mathrm{out}}=\sum_{i,j}\chi_{\textrm{R}ij}\sigma_i\rho_{\mathrm{in}}\sigma_j^{\dag}, \\
  \end{equation}
where $\rho_{\mathrm{in}}$ and $\rho_{\mathrm{out}}$ denote the input and output polarization density matrices, respectively. To ensure that the reconstructed quantum process remains Hermitian and positive semidefinite, we define a $4\times4$ complex lower triangular matrix $T$, whose elements ${t_i},(i=1,...,16)$ are the real-valued parameters to be optimized. The process matrix $\chi_{\textrm{R}ij}$ is then expressed as:
\begin{equation}
  \chi_{\textrm{R}ij}=\frac{T^{\dag}T}{\trace(T^{\dag}T)}.
\end{equation} 
We employed the maximum likelihood method \cite{tomography} to reconstruct the experimental process matrix by measuring output polarization states (via quantum state tomography) for 
$H$, $V$, $D$, $A$, left-circular ($L$), and right-circular ($R$) polarized input states (eigenstates of the three Pauli operators $\sigma_X, \sigma_Y$, and $\sigma_Z$). 
Due to the observed low insertion loss (0.057 dB), the process matrix is reconstructed as a trace-preserving process. The possible slight polarization-dependent loss ($< 0.057$ dB) is effectively characterized as polarization rotation and decoherence.
In our experiment, the output photons from the router are coupled to a single-mode optical fiber, where unwanted extra polarization rotation is introduced before the polarization measurement. 
Therefore, we estimated the process matrix $\chi_\mathrm{R}$ of the free-space optic router by the correction for the quantum process of the fiber (see Appendix II). 
Figure 4 shows the estimated process matrix $\chi_\mathrm{R}$ for all cases of input-output port combinations. The process of the router is close to an ideal identity operation, which can be expressed as $\chi_\mathrm{i}$: $\chi_{II}= 1$ and all other elements are 0. 
The observed process matrices have the fidelity $F= [\Tr(\sqrt{\sqrt{\chi_{\mathrm{R}}} \chi_\mathrm{i}\sqrt{\chi_{\mathrm{R}}}})]^2 > 99.3$\% for all combinations of input and output ports, showing that the input polarization state is highly maintained over the routing process from arbitrary input to output ports.  
Note that the phase of the horizontal polarization in the output port 2 is shifted by $\pi$ relative to the vertical polarization because of the odd number of reflections; the phase shift can be easily corrected by an additional mirror reflection. 
In our experiment, we used the inverted coordinate of the horizontal polarization at the output port 2 to estimate $\chi_\mathrm{R}$.
The imperfection of the process fidelity is due to the slight misalignment between the orientations of the two RTP crystals in the EOMs. The high-fidelity routing of arbitrarily polarized photons also shows the feasibility of routing a constituent single photon in a multi-photon polarization-entangled state, a resource of quantum networking.

Finally, we demonstrate the routing of polarization-mode N00N entangled states with our developed router. The N00N state is known as a crucial light source for high-precision quantum metrology \cite{JPryde2017, Hong2021} and high-sensitivity quantum imaging \cite{imaging} for surpassing the standard quantum limit. A photonic router compatible with such entangled photons can be applied to the 
all-optical storage of photonic entanglement, time and spatial multiplexing of entanglement generation, and temporal synchronization for multi-photon entanglement. 
In our experiment, we tested the routing of an unheralded two-photon polarization-mode N00N state produced by our collinear SPDC source.
Since the input entangled photons are not heralded and the high-voltage drivers limit the repetition rate to a maximum of 1 MHz, we synchronize the delay generator with pump pulses and prescale the trigger rate for EOMs to 1 MHz. The EOMs are operated with a 60\% duty cycle to ensure a sufficient coincidence count rate ($> 5\times10^3$ counts/s at maximum).
Figure 5 shows the measured two-photon N00N-type interference fringes for the states at the input port 1 and the output ports 1 and 2. 
The interference visibilities observed at the output ports 1 and 2 are $V_2 = 97.30(44)\%$  and $97.27(64)\%$, respectively. Within the photon statistic error, these are in excellent agreement with $V_2 = 96.80(24)\%$ at the input port. The observed high interference visibility shows that the indistinguishability and entanglement of orthogonally polarized photons are highly maintained through the routing process. 
The high visibility also indicates that the generation probability of noise photons, which can disturb photon-number correlation and entanglement, is negligibly small. The non-unity visibility for the input state is due to imperfect group delay compensation between the SPDC twin photons by the KTP crystal. 

\begin{figure}[t!]
  \centering
      \includegraphics[trim={65pt 0pt 60pt 0pt},clip,width=\linewidth]{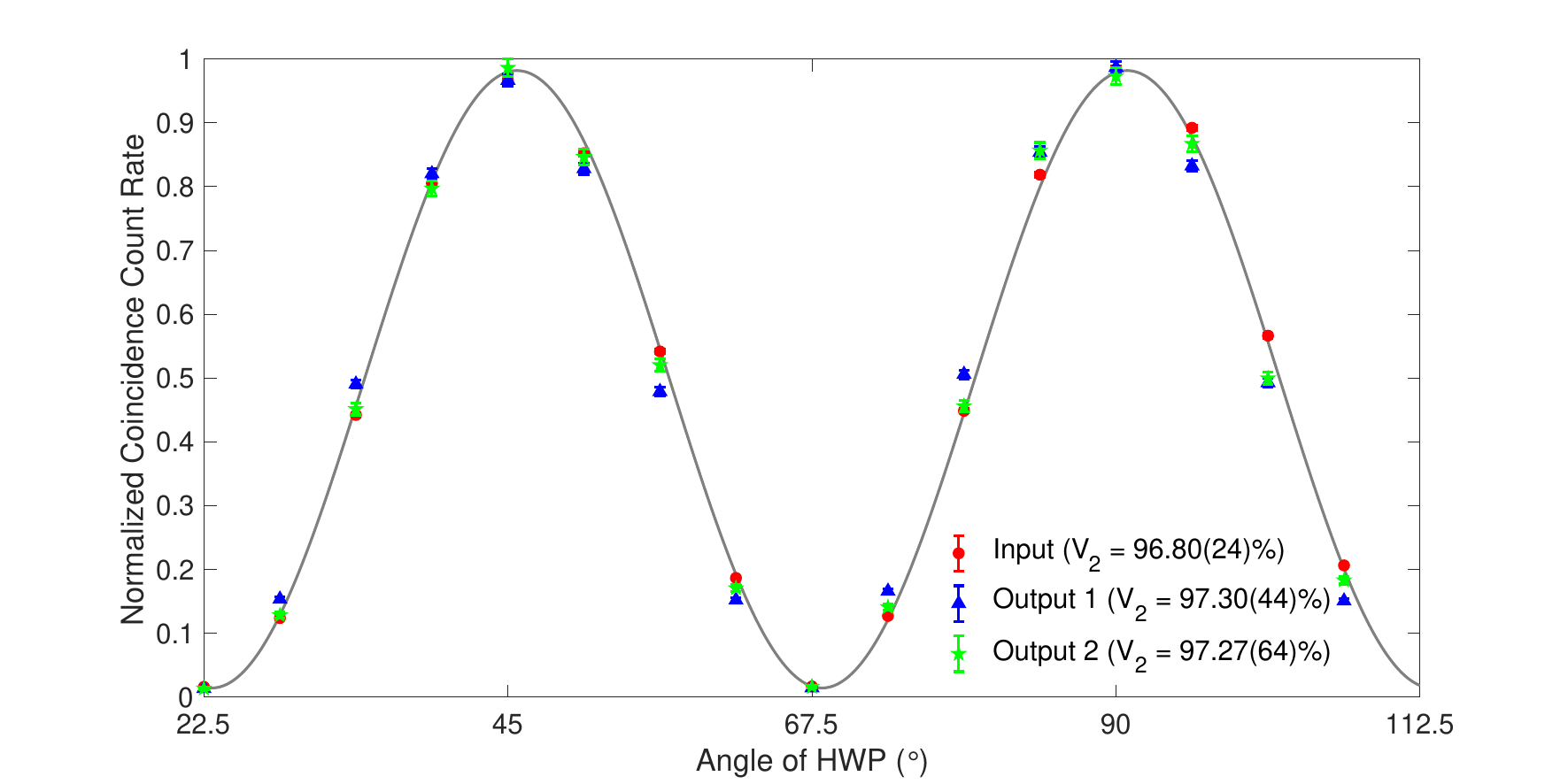}
  \caption{Observed two-photon N00N-type interference at the input port 1 and the output ports 1 and 2. The coincidence count rate between two output ports of PBS is measured while changing the measurement basis with an HWP. The curve shows the sinusoidal fitting for the data points of the input state. 
The interference visibilities at the input and output ports are $V_2 \approx 97\%$ and closely matched with each other. The uncertainties are estimated by Poissonian photon counting statistics.
}
  \label{noonrouting}
  \end{figure}
\begin{table}[ht!]
\centering
\caption{Comparison of performances of polarization-maintaining routers. Free space insertion loss is considered for routers demonstrated in free space. In Ref. \cite{antonZ, LArozema}, the insertion loss was estimated based on a single router setup. Moreover, these works presented the fidelity of routed quantum states relative to their input states, rather than a full quantum process fidelity. To distinguish these values, they are marked with an asterisk (*) for clarification.} 
\begin{tabular}{ p{1.5cm} p{1.7cm} p{1.7cm} p{1cm} p{1.7cm} p{1.3cm} p{1.3cm} } 
\hline
References& Operation wavelength (nm) & Insertion loss (dB) & SER (dB)& Repetition rate (MHz)&Rise(fall) time (ns) & Process fidelity (\%) \\
\hline
\cite{LArozema} & 1550 & $1.3$ & $>20$ & 1 & $< 60$ & $99^*$ \\
\cite{antonZ} &800& $0.13^*$ & 16 & 2.5 & 5.6 & $98^*$ \\ 
\cite{interface} &800& 0.22  & 14.8 & 0.6 & 6 & 94.7 \\ 
\cite{ultrafast}&1310 & 1.30, 1.70 &20&-&$<1$&-\\
\cite{sagnac} &1550& 8.40, 7.60 & 19.21 & $10^3$ & 10 & - \\
\cite{JJAP} &780&0.13&20&0.01&3&$> 99$\\
\hline
This work&1570&0.057&$>22$ & 1 & 3 & $> 99$ \\
\hline
\end{tabular}
\label{comparison}
\end{table}

Table 2 shows the performance of our router in comparison with previous works. Our router shows its superior performance in terms of insertion loss, SER, and polarization process fidelity, owing to our compensator-free configuration with a minimal number of optical components. While previous studies \cite{interface,ultrafast} have demonstrated the routing of a constituent single photon within a two-photon entangled state, our scheme enables the routing of orthogonally polarized multiphoton states with photon-number entanglement, as discussed above. This capability is applicable to quantum applications that require the manipulation of more advanced photonic quantum states, beyond single photons. Despite our characterization performed with the SPDC photons at the L-band, the router is also available at the C-band. In both telecom bands, the low-loss fiber networks, sources of high-indistinguishability single photons \cite{kanedaSA} and high-quality entangled photons \cite{BSM1,BSM2}, and high-efficiency single-photon detectors \cite{SNSPD} have been demonstrated. Although the current loss in the fiber coupling process is 0.41 dB, it can be mitigated to $<$ 0.13 dB by introducing spatial-mode-matching optics \cite{xanadu}.

The stability of the router can also be further improved by miniaturizing the setup and/or utilizing active phase stabilization techniques. 
For achieving even higher stability in interference, a displaced Sagnac interferometer \cite{Okamoto2007} is a widely employed semi-common-path $2\times2$ configuration. The configuration also achieves polarization-maintaining operation with appropriate compensator optics \cite{error-disturbance}. However, to the best of our knowledge, the implementation of such a configuration with a near-normal angle of incidence has not been reported. Further potential improvements to the interferometer configuration will be explored in future work. 
After the feasible improvements, the router will be readily applicable to advanced quantum optics experiments such as storage, synchronization, and multiplexing of entangled photons in the polarization degree of freedom.

\section{Conclusion}
We have experimentally demonstrated a low-loss EO routing of arbitrarily polarized single photons and polarization-based N00N-type entangled photons at the telecom L-band. 
Our scheme is implemented by the low-loss polarization-maintaining EOMs and semi-common-path MZI without compensator optics. All constituent components achieving intrinsic polarization-maintaining operation enable a router configuration with a minimal number of optics, resulting in a low-loss operation for high-efficiency photonic quantum applications. The router experimentally achieved a 0.057 dB insertion loss, a $>$ 22 dB SER, a 3 ns rise time, and a $>$ 99\% polarization process fidelity in the routing of heralded single photons at 1570 nm, where the router can be integrated with low-loss optical fiber cables and high-quality photon sources and detectors. We demonstrated the on-demand routing of the two-photon polarization-mode N00N state for the first time with almost no degradation of its quantum interference visibility. After the feasible improvements in fiber coupling efficiency and interferometer stability, our router can be immediately applicable to all-optical storage of photons for high-efficiency synthesis and measurement of polarization-encoded photonic quantum states \cite{migdall,pittman,multiphoton}. In particular, our router scheme is compatible with time-multiplexing techniques \cite{furusawa,ulrik,entanglemulti} for scaling up quantum information applications utilizing the time domain of optical pulse trains. Low-loss time multiplexing of polarization-encoded qubits and their entanglement with our router scheme will pave the way to large-scale quantum information technologies without chip-scale implementations. Also, the splitting ratio of the router is fully tunable, enabling the usage of universal quantum gates in quantum computing \cite{KLMNature2001,oneway} and quantum metrology \cite{metrology}. We anticipate that our demonstrated scheme will improve various fundamental photonic quantum operations, contributing to the advancement of a wide range of quantum information applications.

\begin{backmatter}
\bmsection{Appendix I. Stability of the router} 
The output stability of our router was characterized using a continuous wave (CW) laser input. Figure 6 shows normalized optical power from the router's two output ports monitored for 5 h. Our semi-common-path design effectively mitigates environmental disturbances to the optical path length difference, enabling stable operation for over 4 h in a laboratory environment. Note that the EOMs were not activated during this specific CW measurement. Our comprehensive characterization experiments utilizing heralded single photons and entangled photons consistently demonstrated the router's stability for more than 2 h, even with the EOMs continuously activated. For example, the total integration time for the rise/fall time data plot (Fig. 6), which spanned approximately 2 h, involved the continuous operation of the EOMs at a 1 MHz repetition rate with the maximum driving voltage applied. This demonstrated stability under active EOM operation indicates the reliability of our router for advanced quantum information applications.
\begin{figure}[t!]
  \centering
      \includegraphics[trim={300pt 205pt 310pt 210pt},clip,width=\linewidth]{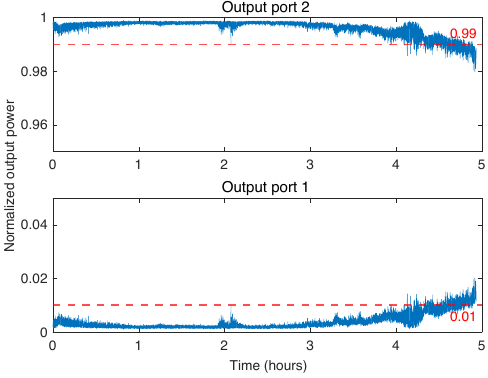}
  \caption{Normalized output power monitored for 5 h. The router maintained output power $> 99\%$ of maximum power for more than 4 h.}
  \label{stability}
  \end{figure}

\bmsection{Appendix II. Estimation of $\chi_\mathrm{R}$}
In our experimental setup, single photons from the router's output ports are subjected to polarization measurement after passing through a single-mode fiber and polarization compensator optics (QWP, HWP, and QWP). However, we found that the compensator optics do not fully compensate for the polarization rotations by the fiber. Thus, the total quantum process $\chi_\mathrm{T}$ estimated from our polarization measurements is indeed the sequential quantum processes of the router $\chi_\mathrm{R}$ and the fiber-to-compensator process $\chi_\mathrm{F}$, which is not an identity operation:
\begin{equation}
  \rho_{\mathrm{out}}=\epsilon_{\mathrm{T}}(\rho_\mathrm{in})=\sum_{i,j}\chi_\mathrm{T}\sigma_i\rho_{\mathrm{in}}\sigma_j^{\dag}=\epsilon_{\mathrm{F}}(\epsilon_{\mathrm{R}}(\rho_\mathrm{in}))=\sum_{m,n}\sum_{i,j}\chi_{\mathrm{F}_{m,n}}\chi_{\mathrm{R}_{i,j}}\sigma_m\sigma_i\rho_\mathrm{in}\sigma_j^{\dag}\sigma_n^{\dag}, \\
\end{equation}
where $\epsilon_{\mathrm{T}}$, $\epsilon_{\mathrm{F}}$, $\epsilon_{\mathrm{R}}$ represent the evolution of $\rho_\mathrm{in}$ by the processes of whole routing system, the fiber and compensator optics, and the free-space router, respectively. 
Quantum process tomography of $\chi_\mathrm{F}$ was performed by placing a set of a high-extinction-ratio polarizer, HWP, and QWP at the output port of the router, which recalibrates and arbitrarily rotates the polarization of single photons. We estimated $\chi_\mathrm{R}$ with obtained tomography results $\chi_\mathrm{T}$ and $\chi_\mathrm{F}$. We used six ($k = \{H, V, D, A, R, L \}$) input polarization states and found $\chi_\mathrm{R}$ by minimizing the cost function:  
\begin{align}
C = \sum_{k} \| \epsilon_{\mathrm{T}}(\rho_{\mathrm{in},k})-\epsilon_{\mathrm{F}}(\epsilon_{\mathrm{R}}(\rho_{\mathrm{in},k})) \|_2,
\end{align}
where $\|A\|_2$ denotes the spectral norm (the largest singular value) of matrix $A$. 
The trace-preserving constraint $\sum_{i,j}\chi_{\textrm{R}ij}\sigma_i\sigma_j^{\dag}=I$ is incorporated into the optimization for a lossless quantum process. Table \ref{rawfidelity} shows the process fidelity of $\chi_\mathrm{T}$, $\chi_\mathrm{F}$, and $\chi_\mathrm{R}$. In our experiment, we employed a single-polarization projection measurement setup. The fiber connections were switched to characterize different output ports, and $\chi_\mathrm{T}$ and $\chi_\mathrm{F}$ were then measured sequentially without disturbing the fiber. The slight differences in the fidelities of $\chi_\mathrm{F}$ at the same output port (e.g., $F_{11} \neq F_{21}$ and $F_{12} \neq F_{22}$) are attributed to the measurement sequence, where $F_{11}$ and $F_{12}$ were measured prior to $F_{21}$ and $F_{22}$. 
The imperfect fiber process, which slightly deviates the photon polarization states (observed as the non-unity fidelity shown in Table \ref{rawfidelity}), is corrected in the estimation of $\chi_\mathrm{R}$ from $\chi_\mathrm{T}$. Consequently, for all input/output cases, we consistently observed the process fidelity of $\chi_\mathrm{R}$ exceeding 99\%. 

\begin{table}[ht!]
\centering
\caption{Process fidelity of $\chi_\mathrm{T}$, $\chi_\mathrm{F}$, and $\chi_\mathrm{R}$. }
\begin{tabular}{ p{1cm} p{1.7cm} p{1.7cm} p{1.7cm} p{1.7cm} } 
\hline
&$F_{11}$ (\%)&$F_{12}$ (\%) &$F_{21}$ (\%) &$F_{22}$ (\%)\\
\hline
$\chi_\mathrm{T}$&98.82(24)&98.76(10)&99.19(16)&99.70(16)\\ 
$\chi_\mathrm{F}$&99.55(18)&98.79(34)&99.73(12)&99.76(8)\\ 
$\chi_\mathrm{R}$&99.56(18)&99.68(22)&99.32(42)&99.61(20)\\
\hline
\end{tabular}
\label{rawfidelity}
\end{table}

\bmsection{Funding}
JSPS KAKENHI (JP21K18902, JP22H01965, and JP25H00847); JST ERATO (JPMJER2402); JST PRESTO (JPMJPR2106).
\bmsection{Disclosures}
The authors declare that they have no conflicts of interest.

\bmsection{Data availability statement}
The data that support the findings of this study are available from the corresponding author upon reasonable request.

\end{backmatter}


\end{document}